\documentclass[a4paper]{jpconf}

\usepackage{graphicx}
\usepackage{amsmath,amssymb}
\usepackage{cite}

\def\bh{black hole}
\def\bhs{black holes}
\def\wh{wormhole}
\def\whs{wormholes}
\def\asflat{asymptotically flat}
\def\bu{black universe}
\def\bus{black universes}
\def\sph{spherically symmetric}
\def\ssph{static, spherically symmetric}
\def\asflat{asymptotically flat}
\def\GR{general relativity}
\def\KS{Kantowski-Sachs}

\def\arXiv{{\it arXiv}:\,}

\def\Jl#1#2{{\it #1} {\bf #2}\ }

\def\CQG#1 {\Jl{Class. Quantum Grav.}{#1}}
\def\GC#1 {\Jl{Grav. Cosmol.}{#1}}
\def\GRG#1 {\Jl{Gen. Rel. Grav.}{#1}}
\def\PRD#1 {\Jl{Phys. Rev. D}{#1}}
\def\PRL#1 {\Jl{Phys. Rev. Lett.}{#1}}

\def\cm{\hspace*{1cm}}

\def\eq{Eq.\,}

\def\al{&\hspace{-.5ex}}
\def\d{\partial}
\def\eq{Eq.\,}

\def\beq{\begin{equation}}
\def\eeq{\end{equation}}               \def\Half{{\dfrac{1}{2}}}
\def\bear{\begin{eqnarray}}            
\def\ear{\end{eqnarray}}               
\def\earn{\nonumber \end{eqnarray}}    \def\kappa{\varkappa}

            \def\diag{\mathop{\rm diag}\nolimits}
      
                    \def\const{{\rm const}}
                \def\eps{\varepsilon}
\def\eql{\al =\al}

\def\R{{\mathbb R}}
\def\mn{_{\mu\nu}}
\def\MN{^{\mu\nu}}
\def\mN{_\mu^\nu}

\begin{document}

\title{Visible, invisible and trapped ghosts as sources
	 of wormholes and black universes}

\author{S V Bolokhov$^1$, K A Bronnikov$^{1,2,3}$,
	 P A Korolyov$^1$\\ and M V Skvortsova$^2$}

\address{$^1$
	Inst. of Gravitation and Cosmology, PFUR,
	6 Miklukho-Maklaya st., Moscow 117198, Russia}
\address{$^2$
	VNIIMS, 46 Ozyornaya st, Moscow 119361, Russia}
\address{$^3$
	National Research Nuclear University ``MEPhI''
	(Moscow Engineering Physics Institute),\\
	Kashirskoe sh. 31, Moscow 115409, Russia}

\ead{kb20@yandex.ru}

\begin{abstract}
We construct explicit examples of globally regular static, spherically
symmetric solutions in general relativity with scalar and electromagnetic
fields, describing traversable wormholes with flat and AdS asymptotics
and regular black holes, in particular, black universes. (A black universe
is a regular black hole with an expanding, asymptotically isotropic
space-time beyond the horizon.) Such objects exist in the presence of
scalar fields with negative kinetic energy (``phantoms'', or ``ghosts''),
which are not observed under usual physical conditions. To account for that,
we consider what we call ``trapped ghosts'' (scalars whose kinetic energy is
only negative in a strong-field region of space-time) and ``invisible
ghosts'', i.e., phantom scalar fields sufficiently rapidly decaying in the
weak-field region. The resulting configurations contain different numbers
of Killing horizons, from zero to four.
\end{abstract}

% --------------------------------------------------------------------

  The so-called exotic matter, which violates the weak and null energy
  conditions (WEC and NEC), is well known to be a necessary ingredient
  for the construction
  of wormholes in general relativity and some of its extensions (see, e.g.,
  \cite{vis-book, lobo-rev, we-book, we-14} for reviews), although such
  matter is not observed under usual physical conditions, and its possible
  existence meets serious theoretical objections mostly related to quantum
  phenomena (see \cite{b-star07} and references therein). On the other hand,
  phantom fields naturally appear in some models of string theory \cite{strings},
  supergravities \cite{sugra} and theories in more than 11 dimensions
  \cite{multi}. One more viewpoint of interest is \cite{h_ellis} that a phantom
  field, being a source of gravity, can exist without its own dynamics and
  therefore escape the related problems with quantum particle creation.

  An important, though not very confident, support from modern cosmological
  observations allowing for the ratio of pressure to energy density
  $w = p/\rho < -1$ (see, e.g., \cite{planck15} and references therein) is
  one of the reasons for the recent interest in possible phenomena with
  exotic matter including \wh\ construction and properties.

  It has also been found that in the presence of exotic matter, say, in the
  form of a phantom scalar field, not only \whs\ are possible but also
  different types of regular black holes, including the so-called \bus.
  The latter look from their static regions as ``ordinary'' \bhs\ in \GR,
  but instead of a singularity beyond the horizon, there is an expanding
  universe which can at large times become isotropic, in particular, de
  Sitter \cite{bu1, bu2}.

  In the present note, we briefly describe different extensions of the scalar
  field solutions of \cite{bu1} to configurations containing electric or
  magnetic fields in the framework of \GR. One of the motivations for
  their inclusion is that by modern observations there can exist a global
  magnetic field up to $10^{-15}$ Gauss, causing correlated orientations of
  sources remote from each other \cite{o-mag}, and some authors admit a
  possible primordial nature of such a magnetic field.

  The most straightforward extension \cite{pha-mag} assumes the action
\beq							\label{S0}
      S = \Half \int \sqrt{-g} d^4 x \Big[
            R + \eps(\d\phi)^2 - 2V(\phi) - F\mn F\MN \Big],
\eeq
  where $R$ is the scalar curvature, $g = \det (g\mn)$, and $F\mn$ is the
  electromagnetic field tensor, and $\eps = \pm 1$ distinguishes usual,
  canonical ($\eps = +1$) and phantom ($\eps=-1$) scalar fields. We consider
  \ssph\ space-times with the general metric
\beq                                                        \label{ds}
      ds^2 = A(u) dt^2 - \frac{du^2}{A(u)} -
      		r^2(u) (d\theta^2 + \sin^2\theta d\varphi^2).
\eeq
  written in the so-called quasiglobal gauge $g_{00} g_{11} = -1$
  convenient for describing Killing horizons which occur at regular
  zeros of the function $A(u)$ \cite{we-book}. Assume that the metric is
  \asflat\ as $u\to\infty$, then a {\it \wh\/} is a geometry
  in which $u \in \R$, $A > 0$ at all $u$, and the metric is \asflat\
  or AdS as $u\to -\infty$. If again $u\in \R$ but at large negative $u$
  the function $A(u) \sim - r^2(u) \to -\infty$ (a de Sitter asymptotic),
  we are dealing with a {\it \bu.}

  In both \wh\ and \bu\ metrics, the area function $r(u)$ has a regular
  minimum at some $u = u_0$. The WEC and NEC are necessarily violated at
  such minima due to the Einstein equations: indeed, one of them reads (see
  \eq (\ref{01}) below) $2 A r''/r = -(T^t_t - T^u_u)$, where $T\mN$ is
  the stress-energy tensor (SET) and the prime stands for $d/du$.
  In terms of the density $\rho$ and radial
  pressure $p_r$, the condition $r'' > 0$ near $u=u_0$ implies
  $\rho + p_r < 0$, which manifests NEC violation. This is true both for an
  R-region ($A > 0$) where $u=u_0$ is a throat and for a T-region ($A < 0$)
  where $u=u_0$ is a bouncing point in a \KS\ cosmology
  \cite{we-book, BK15}.

  The electromagnetic fields can be radial electric (Coulomb) and magnetic
  (monopole) ones. There is, however, no need to introduce specific
  electric or magnetic charges (or monopoles): in both wormholes and black
  universes a radial field can exist without sources due the space-time
  geometry. In the wormhole case it perfectly conforms to Wheeler's idea of
  a ``charge without charge'', in a \bu\ the situation is basically the same
  but looks more involved.

  The electromagnetic field equations are solved in a general form, leading
  to the SET $T\mN [e] = q^2 r^{-4}(u) \diag (1, 1, -1, -1)$, where $q^2
  = q_e^2 + q_m^2$ and the constants $q_e$ and $q_m$ are the effective
  electric and magnetic charges. The remaining independent Einstein equations
  read
\bear
              (A'r^2)' \eql - 2r^2 V + 2q^2/r^2,               \label{00}
\\
                 r''/r \eql - \eps {\phi'}^2,                  \label{01}
\\
         A (r^2)'' - r^2 A'' \eql 2 -4 q^2/r^2.                \label{02}
\ear

  Being interested in \wh\ and \bu\ solutions, we take $\eps=-1$ and,
  using the inverse problem method, assume for the area function
  $r = \sqrt{u^2 + b^2} \equiv b \sqrt{x^2 +1}$, $b = \const > 0$ (the
  length scale). Then \eq (\ref{02}) is integrated giving for
  $B(x) \equiv A(x)/r^2(x)$
\beq                                                       \label{B}
       B(x) = B_0 + \frac{1+ q^2 + px}{1 + x^2}
       + \biggl(p + \frac{2q^2x}{1+x^2}\biggr)\arctan x + q^2\arctan^2 x,
\eeq
  where $p$ and $B_0$ are integration constants. The metric is thus
  completely known while $\phi$ and $V$ are now easily found from the field
  equations. Assuming that our system is \asflat\ at large $x$, so that
  $A\to 1$, $r\to\infty$ and $B\to 0$ as $x\to \infty$, the constants are
  related by
\beq                                                        \label{B_0}
        B_0 = -\pi p/2 - \pi^2 q^2/4, \qquad
        p = 3 m - \pi q^2.
\eeq
  The whole solution is parametrized by two constants $m$ (Schwarzschild
  mass at $x\to \infty$) and $q$, the charge. It turns out that even if we
  restrict ourselves to solutions \asflat\ as $x\to \infty$ and $m >0$,
  there are as many as 10 classes of globally regular space-times which
  differ in the type of the second asymptotic $x\to -\infty$ (M ---
  Minkowski, dS --- de Sitter, AdS --- anti-de Sitter) and the number and
  kinds of horizons, see Table 1 (where $n=1$ refers to simple horizons,
  $n=2$ to extremal ones, R- and T-regions are disposed from left
  to right along the $x$ axis). Two examples of the corresponding
  Carter-Penrose diagrams are shown in Fig.\,1: one of them, with two
  horizons, cannot be drawn on a single plane due to region overlapping
  while the other, with three horizons, occupies the whole plane except
  the grey triangles. Some of non-\asflat\ solutions contain four horizons.
  We refer to \cite{pha-mag} for a detailed description of the global
  causal structures and Carter-Penrose diagrams for different branches
  of the general solution.
\begin{table}[h]
\caption{Types of \asflat\ solutions with $m > 0$.}
\begin{center}
\begin{tabular}{lll}
\br
    Configuration type, asymptotics\ \ \ \ & Horizons:  & disposition of \\
    ($x \to+\infty$) --- ($x\to-\infty$) & number, order $n$ & R- and T-regions\\
\mr
   M -- M \wh & none & R \\
   M -- M extremal \bh & 1 hor., $n=2$ 		& RR \\
   M -- M \bh & 2 hor., $n=1$ (both)   		& RTR \\
   M -- dS \bu & 1 hor., $n=1$ 	       		& TR  \\
   M -- dS \bu & 2 hor., $n=2$ and $n=1$ 	& TTR\\
   M -- dS \bu & 2 hor., $n=1$ and $n=2$ \qquad & TRR\\
   M -- dS \bu & 3 hor., $n=1$ (each)    	& TRTR\\
   M -- AdS \bh & 2 hor., $n=1$ (both)    	& RTR \\
   M -- AdS extremal \bh & 1 hor., $n=2$ 	& RR \\
   M -- AdS \wh          & none          	& R \\
\br
\end{tabular}
\end{center}
\end{table}
\vspace{-6mm}
\begin{figure}[h]
\begin{center}
\raisebox{3.5mm}
{\includegraphics[width=2.5in]{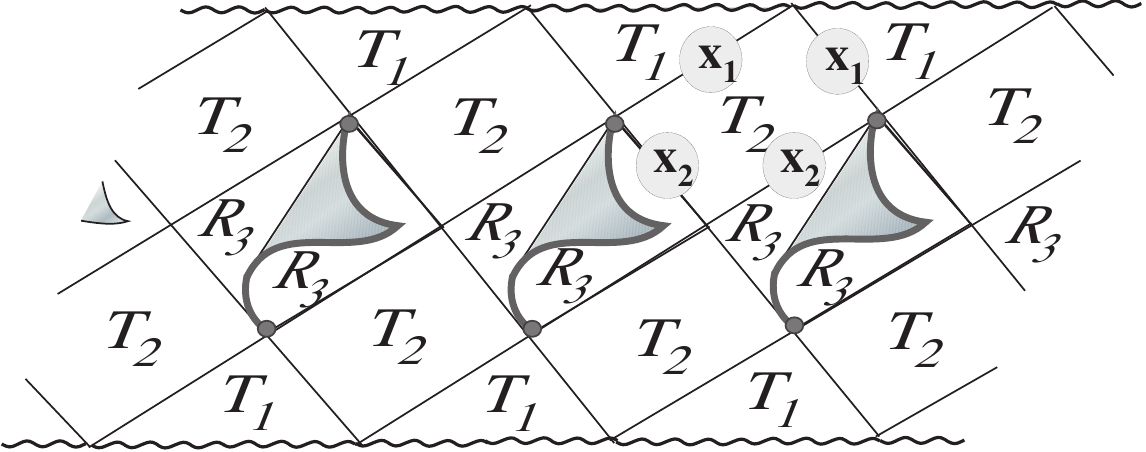}} \cm
\includegraphics[width=1.7in]{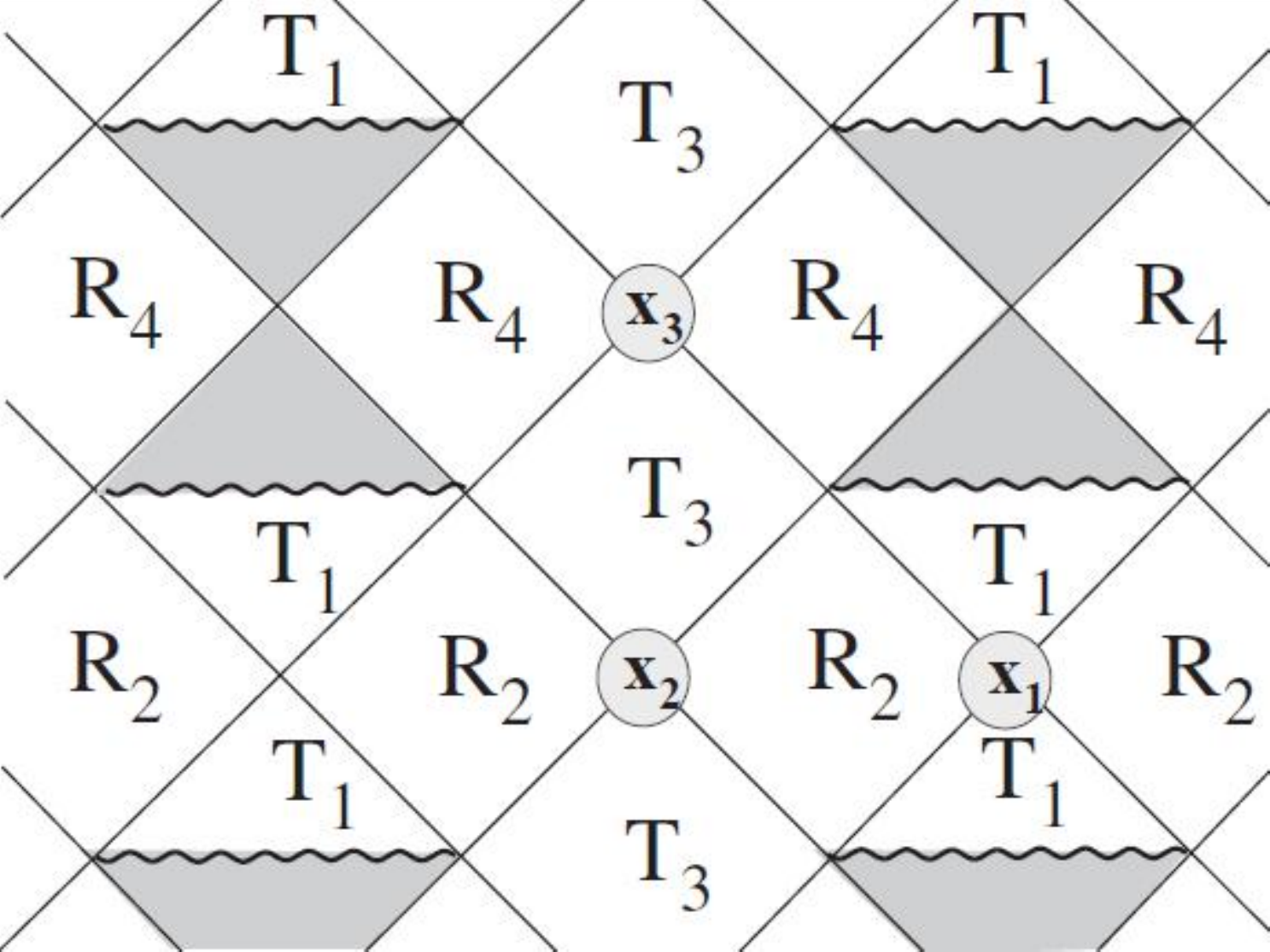}
\caption{Carter-Penrose diagrams of two space-times presented in Table 1,
 	 5th line (left panel) and 7th line (right panel). The
	 letters R and T mark R- and T-regions, $x_n$ mark the
	 horizons, and both are ordered left to right along the $x$ axis.}
\end{center}
\end{figure}

  Since exotic matter or phantom fields are so far not detected, it makes
  sense to try to avoid their emergence in asymptotic weak-field regions.
  To do so, we considered a special kind of fields, named ``trapped ghosts''
  \cite{trap1, trap2, trap3}, which have phantom properties only in some
  restricted strong-field region and become canonical
  outside it. This happens if we substitute in (\ref{S0}),
  instead of $\eps$, a function $h(\phi)$ changing its sign at some $\phi$
  value. Choosing, instead of $r = \sqrt{u^2 + b^2}$,
  the function \cite{trap2, trap3}
\beq                                                            \label{r2}
	  r(u) = a \frac{x^2+1}{\sqrt{x^2+n}}, \qquad
	  a = \const >0, \qquad n = \const > 2.
\eeq
  we obtain examples of such a behaviour. Indeed, with (\ref{r2}), $r''> 0$
  (corresponding to a ghost field) at $x^2 < n(2n\!-\!1)/(n\!-\!2)$ and
  $r''< 0$ at larger $|x|$, as required. As before, all other quantities
  are found from the field equations, and the qualitative features of
  solutions are the same as described above, including the ten classes of
  regular solutions presented in Table 1.

  A weak point of such models is the transition surface from normal to
  phantom fields. It turns out, for example, that if we consider \sph\
  perturbations of solutions with such a field, the corresponding effective
  potential has a singularity which should in general lead to a violent
  instability \cite{BK15}.

  Another approach is to consider what may be called an ``invisible ghost'',
  i.e., a phantom field that decays sufficiently rapidly in the weak-field
  region. In \cite{BK15} this idea was realized with the same ansatz
  (\ref{r2}) on $r(u)$ but with a sigma-model-like combination of two scalar
  fields, a phantom one, $\psi(u)$, and a canonical one, $\phi(u)$, where
  $\psi$ decays at large $u$ much more rapidly than $\phi$. Since the
  function $A(u)$ is, as before, found from \eq (\ref{02}) with known $r(u)$,
  the whole geometry remains the same, changes only the field content of the
  system.

  A configuration with an ``invisible ghost'' can also be obtained with a
  single field $\phi(u)$, but then one should choose a function $r(u)$
  like $r(u) = a(1+ x^8)^{1/8}$, closer than before approaching $r=u$ at
  large $u$ and preserving the condition $r'' > 0$ to ensure the phantom
  nature of $\phi$. The resulting geometries should be qualitatively
  the same as those discussed in \cite{pha-mag, trap3}. Indeed, since
  $r(x)\approx |u|$ at both infinities, the causal structure and the
  corresponding Carter-Penrose diagrams are completely determined by
  zeros of $B(x)$ and its asymptotic behaviour.

  Summarizing, we have obtained numerous examples of \wh, regular \bh\ and
  \bu\ solutions to the Einstein equations with scalar and electromagnetic
  fields as sources, where the scalar fields are phantom in nature at least
  in the strong field region.

\ack
The work of KB was performed within the framework of the Center FRPP supported by
MEPhI Academic Excellence Project (contract No. 02.a03.21.0005, 27.08.2013).


\section*{References}
\begin{thebibliography}{99}     \itemsep 2pt

\bibitem{vis-book}
	Visser M 1995 {\it Lorentzian Wormholes: from Einstein to
	Hawking\/} (Woodbury: AIP)

\bibitem{we-book}
	Bronnikov K A and Rubin S G 2012 \textit{Black Holes, Cosmology
	and Extra Dimansions} (Singapore: World Scientific)

\bibitem{lobo-rev}
	Lobo F S N 2013
%``Time machines and traversable wormholes in modified theories of gravity'',
	{\it EPJ Web Conf.} {\bf 58} 01006 (\arXiv 1212.1006)

\bibitem{we-14}
	Bronnikov K A  and Skvortsova M V 2014
%  Cylindrically and Axially Symmetric Wormholes. Throats in Vacuum?,
	\GC {20} 171 (\arXiv 1404.5750)

\bibitem{b-star07}
	 Bronnikov K A and Starobinsky A A 2007
% No realistic wormholes from ghost-free scalar-tensor phantom dark energy,
     	{\it Pis'ma v ZhETF\/} {\bf 85} 3; {\it JETP Lett.\/} {\bf 85} 1
	(\arXiv gr-qc/0612032)

\bibitem{strings}
	Sen A 2002 {\it JHEP} {\bf 0204} 048; {\bf 0207} 065

\bibitem{sugra}
	Nilles H P 1984 {\it Phys. Rep.} {\bf 110} 1

\bibitem{multi}
	Khviengia N, Khviengia Z, Lu H and Pope C N 1998 \CQG {15} 759

\bibitem{h_ellis}
	Ellis H G 2015 {\it Int. J. Mod. Phys. D} {\bf 24} 1550069

\bibitem{planck15}
	Ade P A R {\it et al\/} (Planck Collaboration) 2015
	Planck 2015 results. XIII. Cosmological parameters \
	\arXiv 1502.01589

\bibitem{bu1}
	Bronnikov K A  and Fabris J C 2006
%	``Regular phantom black	holes'',
	\PRL {96} 251101 (\arXiv gr-qc/0511109)

\bibitem{bu2}
	Bronnikov K A, Melnikov V N and Dehnen H 2007
%	``Regular black	holes and black universes'',
	\GRG {39} 973 (\arXiv gr-qc/0611022)

\bibitem{o-mag}
	Poltis R and Stojkovic D 2010
%	Can primordial magnetic fields seeded by electroweak strings cause
% 	an alignment of quasar axes on cosmological scales?,
	\PRL {105} 161301 (\arXiv 1004.2704)

\bibitem{pha-mag}
	Bolokhov S V, Bronnikov K A and Skvortsova M V 2012
%	``Magnetic black universes and wormholes with a phantom scalar'',
	\CQG {29} 245006 (\arXiv 1208.4619)

\bibitem{BK15}
	Bronnikov K A  and Korolyov P A 2015 \GC {21} 157
	(\arXiv 1503.02956)

\bibitem{trap1}
	Bronnikov K A  and Sushkov S V 2010
%	Trapped ghosts: a new class of wormholes,
	\CQG {27} 095022 (\arXiv 1001.3511)

\bibitem{trap2}
	Bronnikov K A  and Donskoy E V 2011
% 	Black universes with trapped ghosts,
	\GC {17} 176 (\arXiv 1110.6030)

\bibitem{trap3}
	Bronnikov K A, Donskoy E V and Korolyov P A 2013
%	Magnetic wormholes and black universes with trapped ghosts, \
	{\it Vestnik RUDN} No. 2, 139.

\end{thebibliography}
\end{document}